\documentclass[aps, preprint, floats,prb]{revtex4}
\usepackage{graphics,dcolumn}
\usepackage{graphicx}
\usepackage{amsmath}

\begin{document}

\title{Electron transport in real time from first-principles}

\author{Uriel N. Morzan$^{1,*}$}
\author{Francisco F. Ram\'irez$^{1}$}
\author{Mariano C. Gonz\'alez Lebrero$^{1,2}$}
\author{Dami\'an A. Scherlis$^{1,*}$}

\affiliation{$^1$Departamento de Qu\'{i}mica Inorg\'{a}nica, Anal\'{i}tica
y Qu\'{i}mica F\'{i}sica/INQUIMAE, Facultad de Ciencias Exactas
y Naturales, Universidad de Buenos Aires, Ciudad Universitaria,
Pab. II, Buenos Aires (C1428EHA) Argentina}

\affiliation{$^2$Instituto de Qu\'imica y Fisicoqu\'imica Biol\'ogicas, IQUIFIB, CONICET, Argentina}

\email{umorzan@qi.fcen.uba.ar, damian@qi.fcen.uba.ar}

\begin{abstract}

While the vast majority of calculations reported on molecular conductance have been based on the static 
non-equilibrium Green's function formalism combined with density functional theory, in recent years a few
time-depedent approaches to transport have started to emerge. Among these, the driven 
Liouville-von Neumann equation ({\it J. Chem. Phys.} 124, 214708 (2006)) is a simple and appealing route
relying on a tunable rate parameter,
which has been explored in the context of semi-empirical methods. In the present study, we
adapt this formulation to a density functional theory framework and analyze its performance.
In particular, it is implemented in an efficient all-electron DFT code with Gaussian basis functions, 
suitable for quantum-dynamics simulations of large molecular systems.
At variance with the case of the tight-binding calculations reported in the literature,
we find that now the initial perturbation to drive the system out of equilibrium
plays a fundamental role in the stability of the electron
dynamics, and that the equation of motion used in previous tight-binding implementations has
to be modified to
conserve the total number of particles during time propagation.
Moreover, we propose a procedure to get rid of the dependence of the current-voltage
curves on the rate parameter.
This method is employed to obtain the current-voltage characteristic of saturated
and unsaturated hydrocarbons of different lenghts,
with very promising prospects. 

\end{abstract}

\date{\today}

\maketitle

\section{Introduction}

Electron transport  through molecules and nanostructures has been a field of very active research in the last decades, greatly motivated by the interest in molecular electronics and reinvigorated by the often intriguing lack of agreement between calculations and experiments.\cite{Review-Ni-Ra,PRL-Di-Ventra,C4CS00203B,PCCP_2015}
Most of the theoretical approaches currently available are based on the Landauer steady state formalism, formulated in terms of the non-equilibrium Green's function (NEGF) for coherent transport.\cite{Datta} In this context, a usual approximation consists in obtaining the Green function of the system from the  Kohn-Sham (KS) single-particle Hamiltonian ground state. The exchange-correlation potential is approximated by the one used in time-independent density functional theory (DFT), and the charge density is calculated self-consistently in the presence of a steady state current. The effects of the leads attached to the system are represented through the corresponding self-energies.\cite{PRB-Niehaus-GHC,PRB-Transiesta,Datta} This scheme has proved useful to estimate conductance in a variety of molecules and nanoscale structures coupled to semi-infinite leads.\cite{PNAS-Ratner,JACS-Ratner} An important limitation of these calculations is the fact that the transmission function from static DFT has resonances at the non-interacting Kohn Sham excitation energies, which more often than not disagree with the true values.   

Several developments in open or periodic boundary conditions have gone beyond the static picture. Stefanucci and other authors derived rigorous treatments within time-dependent DFT (TDDFT) for the explicit temporal evolution of the system's wavefunction or density matrix, based on the time-dependent Green's function.\cite{Europhys-Stefanucci,PRB-Stefanucci,PRB-2007-GHC,PRB-2005-Kurt-Stefanucci}
Also within DFT, Burke, Car and Gebauer avoided the explicit treatment of semi-infinite leads by using ring boundary conditions.\cite{Gebauer-CPC,PRL-Gebauer-Burke} All these are elegant, though computationally onerous, routes to non-equilibrium transport properties. Presently, the cost of the computations  circumscribes their application to relatively simple models.

On the other hand, the microcanonical dynamics proposed by Di Ventra and Todorov,\cite{DiVentra-Todorov-Microcanonical} readdressed and implemented in a different setting by Cheng and co-workers,\cite{prb_vanvoorhis} is an interesting alternative to the methodologies mentioned above. In this treatment the open-boundary conditions are substituted by a closed set of equations of motion in a finite model, where the leads must be large enough to mimic the discharge in the grand-canonical framework. The initial density for the propagation is taken from a standard DFT calculation in the presence of a bias, which is relaxed at time zero to allow the current to flow from regions of high to low potential. Di Ventra and his collaborators showed how a formally exact current between the leads is established in an ``instantaneous'' or quasi steady state regime. This approach removes the need to implement demanding scattering boundary conditions, but in exchange the size of the leads required to provide reasonable discharge times limits its practical use. Measurements are thus performed in a quasi steady state occurring in a relatively short time window before the electrons are backscattered from the boundaries of the finite leads.

Somehow in between these two general quantum-dynamics frameworks---the microcanonical and the grand canonical ones---, the open-boundary scheme proposed by Sanchez and co-workers is an appealing and conceptually simple method, in which the standard Liouville-Von Neumann expression for the time derivative of the density matrix is augmented with a driving term.\cite{dlvn} This term, which depends on a driving rate parameter ($\Gamma$), allows to maintain the charge imbalance after the external potential is turned off, by restoring the elements of the density matrix associated with the leads back to the polarized state. With this strategy the backscattering inherent to microcanonical dynamics is avoided and the system can reach a steady state. Using a tight-binding model, it has been shown that this method reproduces quantitatively the result obtained with the static Landauer approach.\cite{dlvn} Moreover, in cases where static methods yield multiple current values for a given bias, this dynamical approach is capable of selecting the most stable solution. Subotnik and co-authors have further explored the role and physical meaning of $\Gamma$, replacing the explicit representation of the leads by bath reservoirs where electrons follow an equilibrium Fermi-Dirac distribution.\cite{subotnik} More recently, Hod and collaborators implemented a modified form of the equation of motion which led to improvements in the stability and steady-state convergence of the quantum dynamics.\cite{odedhod} In particular, they introduced a unitary transformation from the orthogonal, tight-binding atomic orbital basis, to a state representation where the new basis elements can be identified with the source, drain, or device. This state representation redefines the bias voltage in terms of the coupling between the eigenstates of the isolated sections of the full system. 

Our goal in the present study is to realize a first-principles implementation of the driven Liouville-von Neumann equation discussed in the previous paragraph, suited for transport simulations in realistic molecular systems. We find that when this equation of motion, either in its original or in its modified form, is integrated in a Kohn-Sham setting with Gaussian basis functions, several issues arise which render it inapplicable to transport calculations. Namely, the trace of the density matrix is not preserved, leading to fluctuations in the total number of particles throughout the propagation, and the steady state current becomes strongly dependent of the $\Gamma$ value, which remains an arbitrary parameter. In this article we introduce a scheme that circumvents these and other flaws observed when the driven Liouville-von Neumann approach is ported to the realm of first-principles simulations.
The method is  implemented in an efficient real time TDDFT code developed in our group, designed for computations in graphic processing units (GPU).\cite{tddft} The result is a powerful  methodology free of adjustable parameters, to accede to time-dependent electron transport properties of large molecular structures. The method is illustrated through its application to hydrocarbon polymers.

\section{Driven Liouville-von Neumann approach}

In tight-binding implementations of the driven Liouville-von Neumann equation,\cite{dlvn,odedhod} the system is divided in three regions: Source, Drain and Molecule (S,D and M respectively). In this context, within an atomistic or site representation for a two lead set up, the density matrix and Hamiltonian can be written as in equations 1 and 2 respectively.
\begin{equation}
 \hat{\rho} = 
\begin{bmatrix}
 \hat{\rho_{S}} & \hat{\rho}_{SM}  & \hat{\rho}_{SD} \\
 \hat{\rho}_{MS}  & \hat{\rho}_{M}  & \hat{\rho}_{MD}  \\
  \hat{\rho}_{DS}  & \hat{\rho}_{DM}  & \hat{\rho}_D  \\
 \end{bmatrix}
\end{equation}

\begin{equation}
 \hat{\bf H} = 
\begin{bmatrix}
 \hat{H}_{S} & \hat{H}_{SM}  & \hat{H}_{SD} \\
 \hat{H}_{MS}  & \hat{H}_{M}  & \hat{H}_{MD}  \\
  \hat{H}_{DS}  & \hat{H}_{DM}  & \hat{H}_D  \\
 \end{bmatrix}
\end{equation}
The quantum-dynamics originally proposed by S\'anchez and co-authors,\cite{dlvn} departs from a ground state density obtained in the presence of an electric field in the Source-Drain direction, which is turned off during the time propagation. In order to avoid the backscattering of the electrons at the boundaries, while keeping a voltage imbalance between the leads, open boundary conditions are introduced by augmenting the standard Liouville Von Neumman equation of motion for the density matrix with a driving term:
\begin{equation}
 \frac{\partial {\bf \hat {\rho}}}{\partial t} = -\frac{i}{\hbar} [{\hat{\bf {H}}}, {\hat{\bf {\rho}}}]  
- \Gamma ({\hat{\bf {\rho}}}-{\hat{\bf {\rho}}}^0)
\end{equation}
where $\Gamma$ is the driving rate parameter and ${\hat{\bf {\rho}}}^0$ can be defined as follows:
\begin{equation}
  {{\bf {\rho}}}_{ij}^0 = \left\{
        \begin{array}{ll}
            {{\bf {\rho}}}_{ij}(t_0) & \text{if  $i,j$ $\in$ $S \cup D$ } \\
            {{\bf {\rho}}}_{ij}(t) & \text{if $i,j$ $\notin$ $S \cup D$ }\\
        \end{array}
    \right.
\end{equation}
Thus, the second term on the right hand side of equation 3 continuously drives the charge in the leads region towards the initially polarized state, but does not directly affect the evolution of the density in the central Molecule region in between.
The two contributions to the driving term, $-\Gamma \hat{\bf{\rho}}$ and $\Gamma \hat{\bf{\rho}^0}$, can be identified with  electron absorption and injection, respectively. We note that absorption and injection processes take place simultaneously, both in Source and Drain. It is the balance between these two contributions what defines the net amount of electrons that will be injected/absorbed in each electrode, and therefore the overall current flowing between them. Starting from the formulation above, Hod and co-workers\cite{odedhod} proposed a modified working expression  in which the damping contribution affects not only the pure lead elements $S$, $D$, $SD$ and $DS$, but also those coupling the leads and the molecule: $SM$, $MS$, $MD$ and $DM$:
\begin{equation}
 \frac{\partial {\bf \hat {\rho}}}{\partial t} = -i [\hat{\bf H}, \hat{\bf \rho}]
-\Gamma
\begin{bmatrix}
 \hat{\rho}_S -  \hat{\rho}_{S}^0 &  \frac{1}{2} \hat{\rho}_{SM}  & \hat{\rho}_{SD} \\
\frac{1}{2} \hat{\rho}_{MS}  & \hat
{0} & \frac{1}{2} \hat{\rho}_{MD}  \\
\hat{\rho}_{DS} & \frac{1}{2} \hat{\rho}_{DM}  & \hat{\rho}_D - \hat{\rho}_{D}^0
 \end{bmatrix}
\label{eqnhod}
\end{equation}
It was shown in the same study that this expression can be derived from the formalism of complex absorbing potentials, in which context the addition of an imaginary potential to the Hamiltonian provokes a damping of the wavefunction and therefore a depletion of electronic density.\cite{odedhod} The modification of the standard Liouville-von Neumann equation with imaginary absorbing potentials of magnitude $\Gamma$ in the Source and Drain regions, leads to a damping term as appearing in equation \ref{eqnhod}.
The final form of this equation is recovered if electron injection is represented in an analogous way, by including a potential of the same magnitude but opposite sign acting on the initial charge of the lead regions, $\hat{\rho}^0_S$ and $\hat{\rho}^0_D$. This ensures that the injected electrons have the equilibrium distribution of the leads subject to the external bias, assuming that deep inside the semi-infinite Source and Drain, the electronic structure remains unperturbed. Using a tight-binding model, the authors found that expression \ref{eqnhod}
yielded an improved dynamics, preserving state occupations and density matrix positivity, 
accelerating the convergence to a steady state and reducing at the same time the noise 
in the current.\cite{odedhod}

\section{First-principles formulation}

Equation \ref{eqnhod} was implemented in an all-electron, Gaussian basis-sets
DFT code developed in our group.\cite{jctc-nano,tddft}
On the basis of GPU parallelization
of the most demanding parts of the computation---which include
the exchange correlation energy and the commutators between $\hat H$ and $\hat \rho$---and
other algorithmic optimizations, this scheme can handle time-dependent simulations
of molecular systems above a hundred atoms, propagated for several hundreds of femtoseconds.\cite{tddft}
The present calculations were performed using 6-31G** basis sets in combination with the
PBE exchange-correlation functional, with a time-step of 0.1 a.u. to integrate the
equation of motion through the Magnus expansion.\cite{tddft}

\begin{figure}
\begin{center}
\includegraphics[width=12cm]{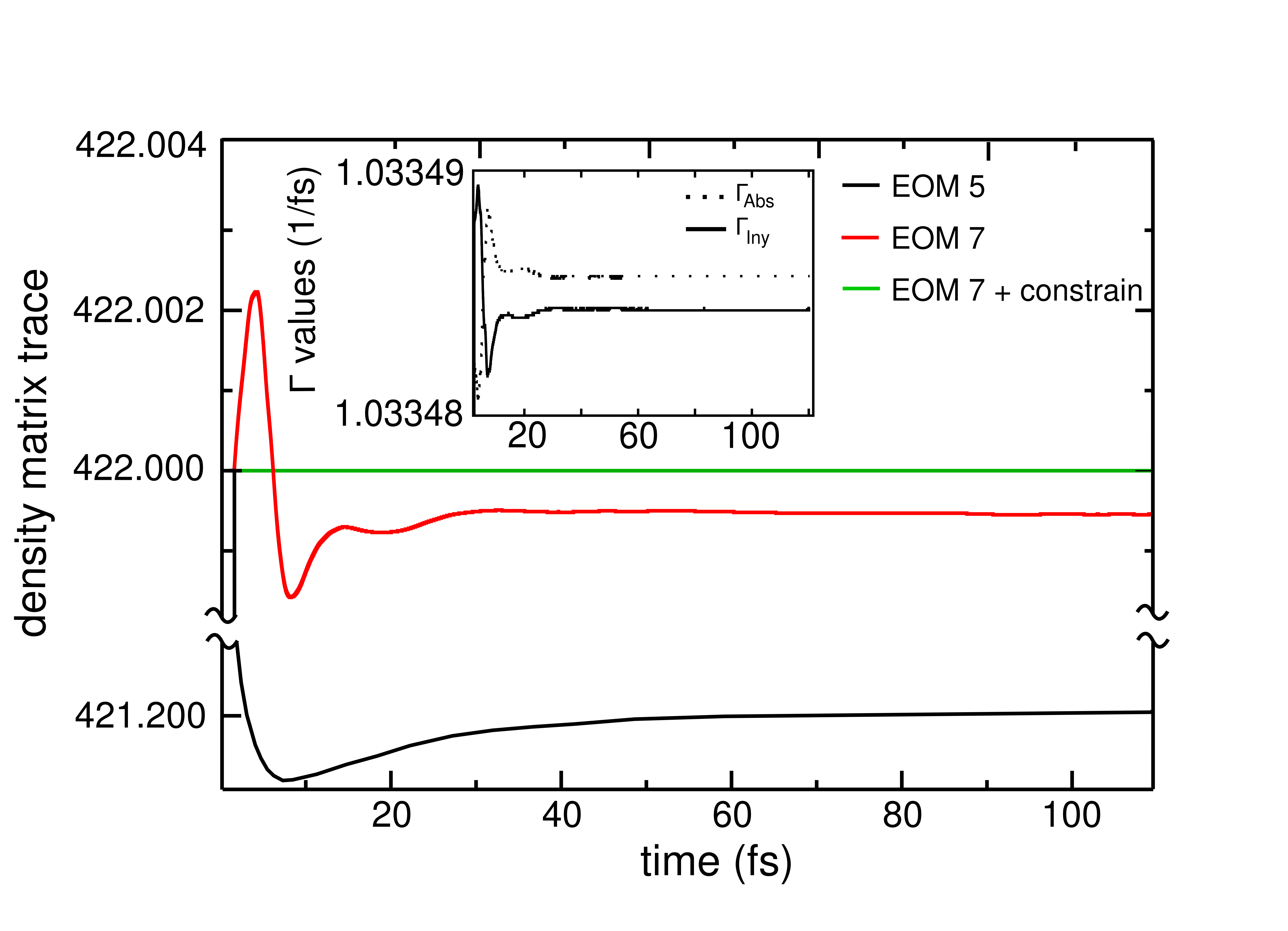}
\end{center}
\caption{Time evolution of the total number of electrons for a trans-polyacetylene chain of
60 carbon atoms. The black, red, and green curves correspond to the results obtained
respectively from equations of
motion \ref{eqnhod}, \ref{mastereqn0}, and \ref{mastereqn} with the constraint on the trace of the
driving term. Inset: absorbing (dotted line) and injecting (solid line) driving rate parameters
as a function of time (fs), for the simulation represented in the main graph with a green line.}
\label{tracevstime}
\end{figure}

The behavior of the working formula \ref{eqnhod} inserted in this density functional scheme,
is illustrated by the dotted curve in Figure \ref{tracevstime} 
for the case of a trans-polyacetylene chain of 60 carbon atoms (CH$_2$-(CH)$_{58}$-CH$_2$), 
where the Source, Molecule, and
Drain fragments consist of 20 carbon atoms each.
The total number of electrons is not conserved during the dynamics, 
but experiences a rapid decrease along the first 7 fs, 
and then slowly stabilizes around a value 0.8 $e$ below the initial charge.
The reason for this unbalance may be tracked in the fact that, 
despite the treatment for injection and absorption in expression \ref{eqnhod} is essentially identical, 
the non-diagonal blocks ($MS$, $SM$, $DM$ and $MD$) in the equation of motion have only absorbing contributions. For sufficiently large lead models, affordable in tight-binding calculations, the impact of these non-diagonal blocks to the overall absorption-injection process may be negligible. However, in first-principles implementations where the computational burden limits the size of the leads, the contribution of  non-diagonal elements to the transport process may become important. As a matter of fact, the incorporation of charge injection in the off-diagonal elements according to equation \ref{mastereqn0}, significantly improves the absorption-injection balance in our TDDFT simulations. This effect is visible in Figure \ref{tracevstime}, which shows that the drift in the total number of particles associated with expression \ref{eqnhod}, is largely eliminated when it is replaced by equation \ref{mastereqn0}. 
\begin{equation}
 \frac{\partial {\bf \hat {\rho}}}{\partial t} = -i [\hat{\bf H}, \hat{\bf \rho}]
-\Gamma
\begin{bmatrix}
 \hat{\rho}_S -  \hat{\rho}_{S}^0 &  \frac{1}{2} (\hat{\rho}_{SM} -  \hat{\rho}_{SM}^0) & \hat{\rho}_{SD} -  \hat{\rho}_{SD}^0 \\
\frac{1}{2} (\hat{\rho}_{MS} - \hat{\rho}_{MS}^0)  & \hat
{0} & \frac{1}{2} (\hat{\rho}_{MD} - \hat{\rho}_{MD}^0)  \\
\hat{\rho}_{DS} -  \hat{\rho}_{DS}^0 & \frac{1}{2} (\hat{\rho}_{DM} -  \hat{\rho}_{DM}^0) & \hat{\rho}_D - \hat{\rho}_{D}^0
 \end{bmatrix}
\label{mastereqn0}
\end{equation}

Strictly, the Hamiltonian and density matrix are defined for a fixed number of particles. Any departure from the initial number along the time-propagation would violate Pauli's exclusion principle, 
leading to unphysical states.
A possible way to suppress these fluctuations altogether, is through the implementation of distinct, time-dependent absorbing and injecting rates, $\Gamma_A (t)$ and $\Gamma_I (t)$, which may be allowed to vary to preserve the total charge in the system. Hence, equation \ref{mastereqn0} assumes the following form:
\begin{equation}
 \frac{\partial {\bf \hat {\rho}}}{\partial t} = -i [\hat{\bf H}, \hat{\bf \rho}] -
\begin{bmatrix}
 \Gamma_A \cdot \hat{\rho}_S -  \Gamma_I \cdot \hat{\rho}_{S}^0 &  \frac{1}{2} (\Gamma_A \cdot \hat{\rho}_{SM} -  \Gamma_I \cdot \hat{\rho}_{SM}^0) & \Gamma_A \cdot \hat{\rho}_{SD} -  \Gamma_I \cdot \hat{\rho}_{SD}^0 \\
\frac{1}{2} (\Gamma_A \cdot \hat{\rho}_{MS} - \Gamma_I \cdot \hat{\rho}_{MS}^0)  & \hat
{0} & \frac{1}{2} (\Gamma_A \cdot \hat{\rho}_{MD} - \Gamma_I \cdot \hat{\rho}_{MD}^0)  \\
\Gamma_A \cdot \hat{\rho}_{DS} - \Gamma_I \cdot \hat{\rho}_{DS}^0 & \frac{1}{2} (\Gamma_A \cdot \hat{\rho}_{DM} - \Gamma_I \cdot \hat{\rho}_{DM}^0) & \Gamma_A \cdot \hat{\rho}_D - \Gamma_I \cdot \hat{\rho}_{D}^0
 \end{bmatrix} ,
\label{mastereqn}
\end{equation}
whereas charge conservation can be enforced by setting to zero the total number of particles associated with the driving term, which can be computed as the trace of the product between the two matrices $\hat{D}$ (the driving operator, equal to the second term on the right hand side of equation \ref{mastereqn}) and $\hat{S}$ (the overlap matrix of non-orthogonal basis functions, $S_{ij} = \langle \phi_i | \phi_j \rangle$):
\begin{equation}
    \text{tr} (\hat{D} \hat{S}) = \sum_{\alpha, \beta} D_{\alpha \beta} S_{\beta \alpha} = \Gamma_A \sum_{\alpha, \beta \notin \text{M,M}} 
\rho_{\alpha \beta} S_{\beta \alpha} - \Gamma_I \sum_{\alpha, \beta \notin \text{M,M}} \rho^0_{\alpha \beta} S_{\beta \alpha}  = 0
    \label{traza}
\end{equation}
The later equation provides the ratio between $\Gamma_A (t)$ and $\Gamma_I (t)$ necessary to keep constant the
number of electrons. Yet, to univocally determine the values of the rate parameters, a constraint is imposed to their sum,
\begin{equation}
    \Gamma_A(t) + \Gamma_I(t) = 2 \Gamma = \text{constant}
\label{suma}
\end{equation}
In this way, $\Gamma_A (t)$ and $\Gamma_I (t)$ are recomputed at every time-step of the dynamics to satisfy simultaneously \ref{traza} and \ref{suma},
which supresses any deviations in the number of electrons (see solid line in Figure \ref{tracevstime}).
Condition \ref{suma} is somehow arbitrary, and aims at keeping the values of the rate parameters close to the constant $\Gamma$.
Other artifacts are also possible, for example to fix one of the two parameters, calculating the other from equation \ref{traza}.
In any case, as can be seen in Figure \ref{tracevstime}, the constraints in equations \ref{traza} and \ref{suma} have in practice a negligible impact
on the transport process, because $\Gamma_I(t)$ and $\Gamma_A(t)$ remain essentially equal and constant throughout the dynamics.

\begin{figure}
\centering
\includegraphics[width=12cm]{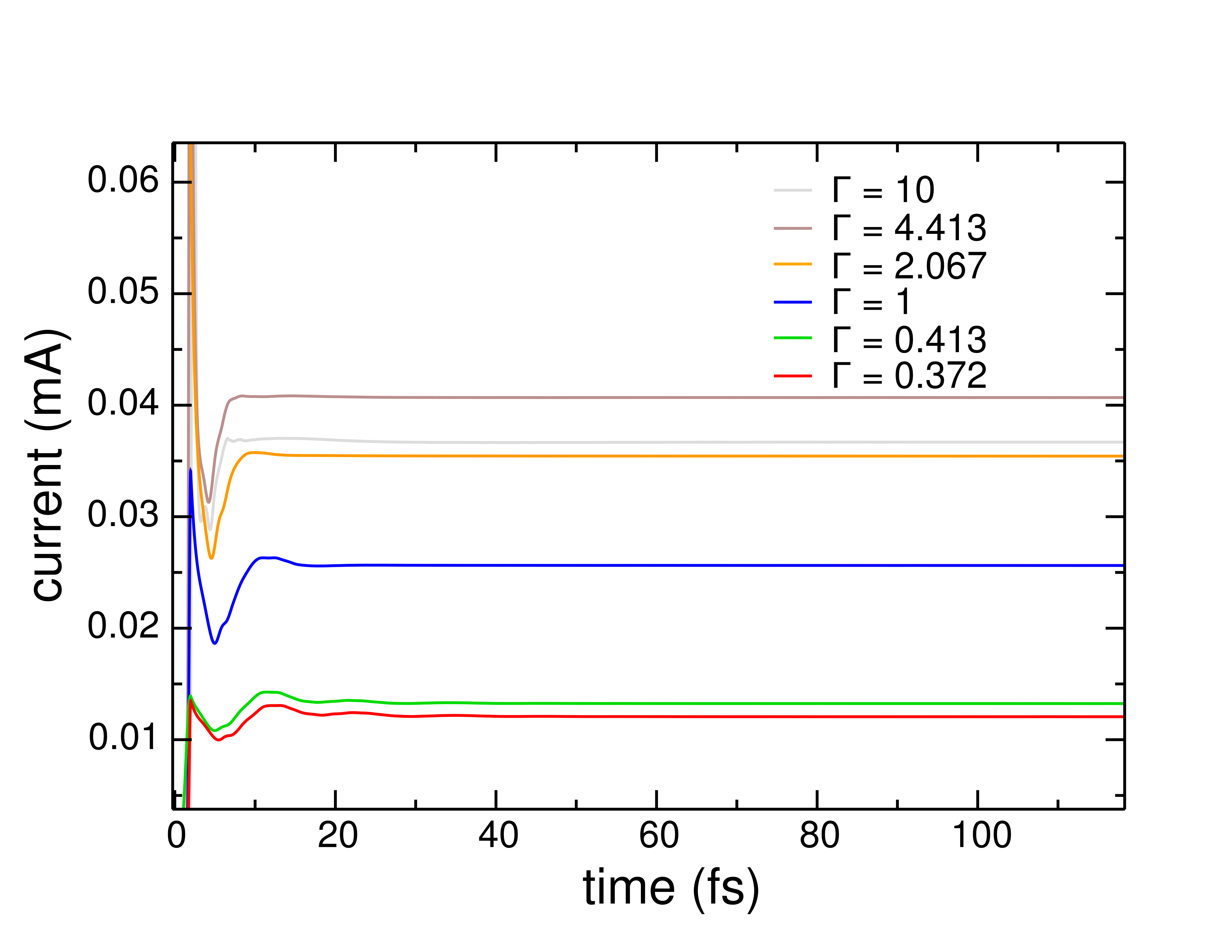}
\caption{Current as a function of time on a trans-polyacetylene chain of
60 carbon atoms, for different values of the rate parameter $\Gamma$ (in atomic units)
and an applied bias of 2.1 V. }
\label{ivsgamma}
\end{figure}

Expression \ref{mastereqn} will be our master equation along this work, with $\Gamma_A$ and $\Gamma_I$
calculated at every time-step according to relations \ref{traza} and \ref{suma}.
The driving matrix $\hat D$ is computed in a non-orthonormal atomic basis, 
and then transformed to an orthonormal representation following the canonical transformation 
process. Equation of motion \ref{mastereqn} is then integrated in this orthonormal basis using
the Magnus propagator to first order.\cite{tddft}
We emphasize that, despite the fact that in this scheme the number of particles 
in the entire system is fixed by virtue of the constraint on the trace of the
driving operator, the total charge in the Molecule region can change during the quantum
dynamics at the expense of an opposite change in the leads. 
The current $I(t)$ flowing between the electrodes can be directly computed as the charge associated
with the absorbing or the injecting terms, 
\begin{equation}
I(t) = \frac{1}{\Delta t} \Big[ \Gamma_A \sum_{\alpha, \beta \in \text{L,L}} 
\rho_{\alpha \beta} S_{\beta \alpha} - \Gamma_I \sum_{\alpha, \beta \in \text{L,L}} 
\rho^0_{\alpha \beta} S_{\beta \alpha} \Big]
\end{equation}
where $L$ represents either the source or drain lead and $\Delta t$  the time-step.
The current achieved in the steady-state depends on the value of $\Gamma$, as shown in Figure \ref{ivsgamma} 
for a fixed $\hat \rho^0$. In the limit of small $\Gamma$, the standard microcanonical picture is
recovered and the backscattering effect precludes any net current.
The increase of the rate parameter exacerbates both injection and absorption by promoting
electron exchange with the reservoirs. There is a relatively broad interval of $\Gamma$ values
which maximize the conductance, above which the damping prevails and the current starts to decay.
The same dependence with the rate parameter has been observed for semi-empirical hamiltonians
by various authors, including Nitzan,\cite{subotnik} Todorov,\cite{dlvn} and,
to a lesser extent, by Zelovich, Kronik and Hod in their model systems.\cite{odedhod}
While in the present DFT simulations the steady state current appears to be more sensitive to the rate parameter
in comparison to previous tight-binding reports, this dependence 
is in any case a weakness of the driven Liouville-von Neumann approach.
Despite the qualitative insights provided in the literature about the  role of 
the driving rate parameter,\cite{dlvn, odedhod, ML-Coup-Hod-JCTC} there is no rigorous or practical way 
to determine it univocally.
We will come back to this issue in section V, where we discuss a path to obtain
the current as a function of the voltage bias, which gets rid of the $\Gamma$ dependence. 

\section{Perturbing the system out of equilibrium}
\begin{figure}
\begin{center}
\includegraphics[width=10cm]{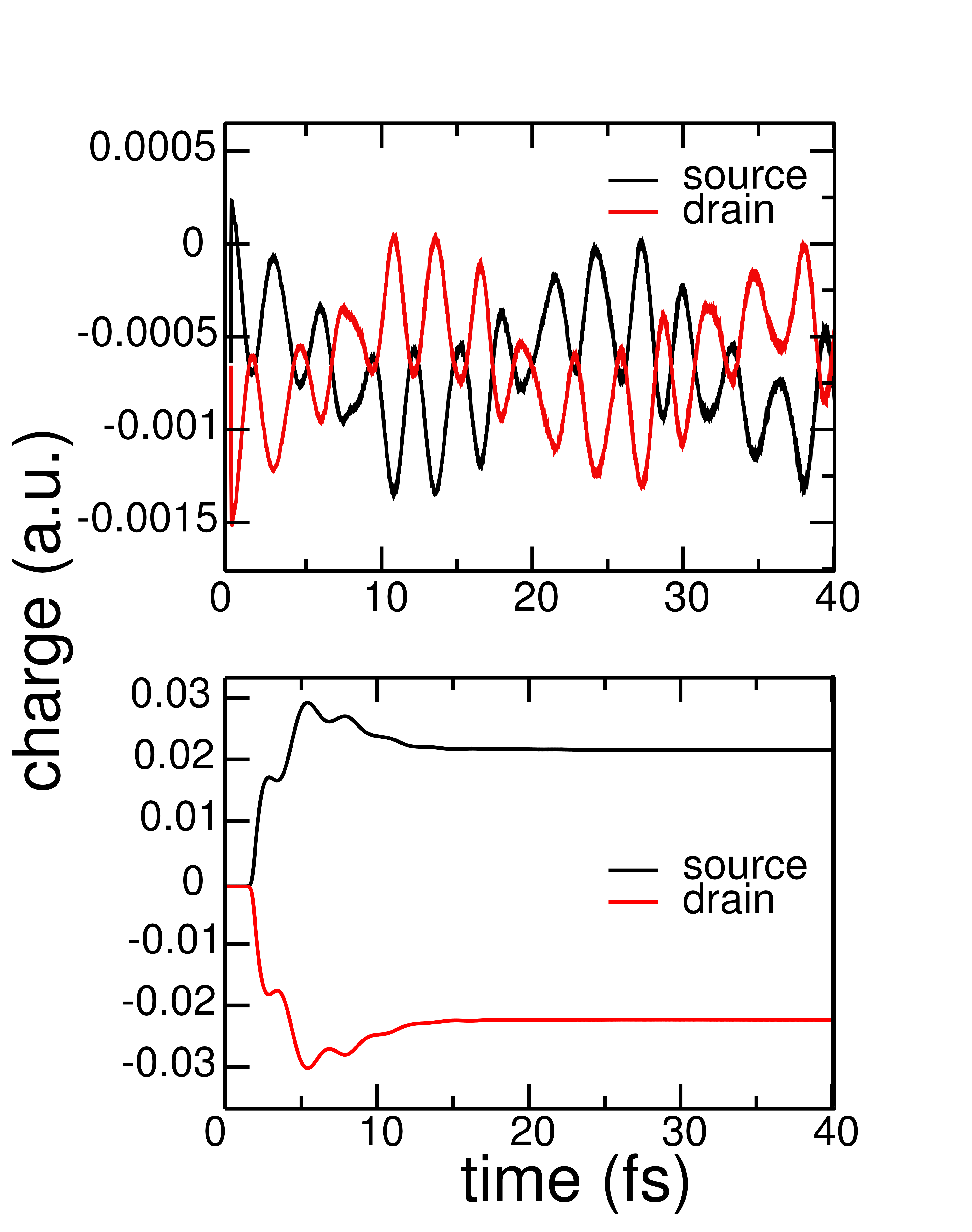}
\end{center}
\caption{Time evolution of the Mulliken charges in the Source and Drain regions, for a
driven dynamics in a polyacetylene chain of 60 carbon atoms. The upper panel
depicts the behavior when the driving term is switched on at time zero. The lower panel corresponds to
a smooth inclusion of the driving term according to equation \ref{eqsmooth}. }
\label{mulliken}
\end{figure}

In these simulations, the density $\hat \rho^0$ involved in the construction of the driving operator is obtained from
a ground-state self-consistent calculation
in the presence of an applied bias. The initial density $\hat{\rho}(t_0)$, instead, and at variance with the practice adopted
in reference \cite{dlvn} , corresponds to the ground-state in the absence of any external field. While it could be also possible
to set $\hat{\rho}(t_0) = \hat \rho^0$, there is a reason for  our particular choice which
will be clear below.  The upper panel of Figure \ref{mulliken} displays the time evolution of the Mulliken 
charges in the Source and Drain regions for a dynamics performed according to this procedure in the polyacetylene molecule.
Conversely to the expected behavior, the populations do not reach stable values, neither the charge of the Source
remains above that of the Drain, but they exhibit an oscillatory and overlapping evolution within this simulation time-scale.

The behavior changes dramatically if the effect of the driving term is incorporated smoothly, by multiplying the driving
rate parameter by a time-dependent factor which increases gradually from 0 to 1. In particular,
the lower panel of Figure \ref{mulliken} shows a more meaningful behavior when the value
of $\Gamma$ is controlled in the following form:
\begin{equation}
  \Gamma = \left\{
        \begin{array}{ll}
            \Gamma_0 \cdot e^{-(t-b)^2/c} & \text{if  $t \leq b$ } \\
            \Gamma_0 & \text{if $t > b$ }\\
        \end{array}
    \right.
 \label{eqsmooth}
\end{equation}
with the chosen numerical parameters $b=2.419$ fs and $c=0.585$ fs$^2$. It was checked that the specific values of $b$ and $c$ have practically no
effect on the final charges and currents in the steady-state. Now, the populations of the leads reach a steady-state in which 
the charge difference between Source and Drain has the expected sign. 

To rationalize these results, it must be recalled that the electron dynamics evolves according to equation \ref{mastereqn} in the absence of an electric field.
Since the starting density corresponds to the ground-state of $\hat H$,
the value of the commutator $[\hat H,\hat \rho]$ is initially zero. However,
the incorporation of the driving term may produce abrupt variations of the density at the initial stages of the time-propagation,
when $\hat \rho - \hat \rho^0$ can be large. The irruption of the perturbation
seemingly excites the accessible resonances of the electron structure, ressembling the
application of a step or delta-function potential in a quantum-dynamics.
Thus, the resulting pattern reflects the characteristic frequencies of the system,
rather than the transport process itself.

The perturbing effect of the driving term could be minimized if the initial density were set
equal to $\rho^0$. However, in such a case $\hat H$ and $\hat \rho$ would not commute,
and similar excitations will develop. With the use of a starting density that is a
solution of $\hat H$, a gradual perturbation is easy to implement through the driving
rate parameter, while its implementation would not be as straightforward if $\hat \rho(t_0) = \hat \rho^0$.
Noteworthy, previous TDDFT studies of molecular conductance using microcanonical dynamics
have reported ``transient fluctuations'' or ``noise'' in current and charges,
which origin could not be established with certainty.\cite{prb_vanvoorhis, NL-VanVoorhis} The results presented above
suggest that the fluctuating nature of those observables might have been a consequence of the sudden 
incorporation or removal of the bias potential in the time dependent Hamiltonian at time zero.
Hence, whereas the initial magnitude of the perturbation does not seem to be an issue for tight-binding models,
it appears to be a crucial ingredient in real-time conductance simulations from first-principles.

\section{The bias potential: dynamical approximation}

Up to this point, the bias potential or voltage bias ($V$) has not been given explicitly,
but has remained implicit in $\hat \rho^0$, which is the ground-state density 
in the presence of a uniform electric field of magnitude $V/d$, 
with $d$ the separation between Source and Drain. 
Equation \ref{mastereqn} continuously drives 
the charge in the reservoirs towards the reference density of the system equilibrated with
the electric field turned on. Nevertheless, in our simulations the electronic density in the
leads, $\hat \rho(t)$, never becomes equal to $\hat \rho^0$---the difference
between the two strongly depends on $\Gamma$---and therefore it would be inaccurate
to assume that the bias that led to $\hat \rho^0$ in a static calculation is the same as the one
developed in dynamical conditions with a driving operator formed with $\hat \rho^0$. 
Hereafter, we will refer to these two magnitudes
as the static ($V_s$) and dynamical ($V_d$) bias. The former is an artifact to
generate $\hat \rho^0$ and construct the driving term, while the later is the effective 
or physical electric
potential difference arising between Source and Drain during the electron dynamics, 
and is the relevant parameter to characterize conductance.
Its value, however, is not known a priori. 
\begin{figure}
\begin{center}
\includegraphics[width=15cm]{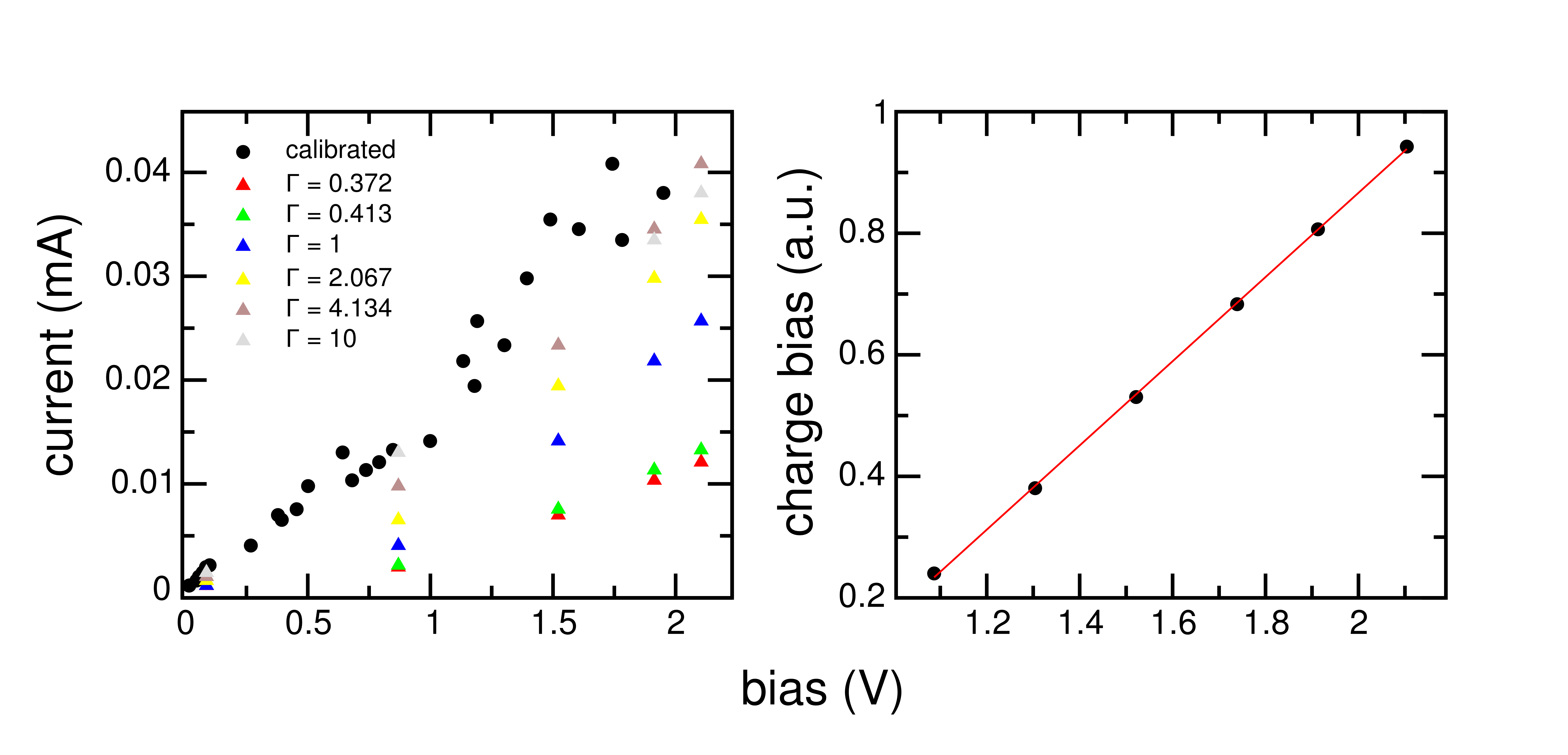}
\end{center}
\caption{Steady-state currents in polyacetylene and the calibration strategy.
On the left panel, the triangles show the steady-state currents as a function of
the statically applied bias ($V_s$)
for different values of the rate parameter $\Gamma$, in atomic units.
The black circles correspond to the same data, expressed as a function of the
dynamical bias ($V_d$) which is assigned according to the charge difference between
Source and Drain in the steady-state. Specifically,
$V_d$ is interpolated from the calibration curve shown
on the right panel, which collects the Mulliken charge difference
surging between Source and Drain in a ground state calculation as a function of the
applied bias.}
\label{biasbias}
\end{figure}

An estimation of $V_d$ can be accomplished in terms of the charge populations of the leads in operating
conditions. In particular, it is possible to establish a
link between $V_d$ and $V_s$ through a sort of calibration
curve based on the charge difference between the leads.
To illustrate this, Figure \ref{biasbias} presents on the left panel the
steady state currents as a function of $V_s$, for various values of $\Gamma$.
It can be seen that, at least in the explored range, 
the current increases with $V_s$ (due to the rise in $\hat \rho^0$),
but the $I$-$V$ curve is not univocally determined because, as already discussed in section III, 
the conductance exhibits a significant dependence on the driving rate parameter.
On the right panel, the Mulliken charge difference obtained in a static calculation 
subject to a uniform field of magnitude $V_s/d$,
is plotted against the corresponding voltage bias $V_s$. This plot serves as a calibration curve
from which, for any given charge difference measured in operation conditions, and in
particular in the steady-state,  
it is possible to interpolate a bias. Thus,
the effective or dynamical bias $V_d$ is estimated as the one that in static
conditions produces the same charge population difference as in the steady-state.

The black circles on the left panel of Figure \ref{biasbias} correspond to the same data-points
obtained for all different $\Gamma$,
after  reassigning the bias according to the calibration curve.
With this approach the dependence of the steady state current on the driving rate parameter 
is practically eliminated, providing a criterion to define the voltage unambiguously.
The calibration also yields the allignment of the data-points on a single line for the rest of
the molecules examined in this work, as it is shown in the next section. 
Thus, by selecting the charge difference as the
reference variable, this procedure offers a way to estimate the electric
potential difference in the dynamical regime, neutralizing at the same time the dependence
on the $\Gamma$ parameter.

We note that the calibrated $I$-$V$ curve tends to be coincident with the curves
corresponding to rate parameters which optimize the conductance (e.g., $\Gamma$=4.134 a.u. or
$\Gamma$=10 a.u.). This means that with the use of these parameters, the
voltage bias applied statically to polarize the system, is approximately maintained
during the transport process. Moreover, this is in line with the analysis in reference \cite{odedhod} ,
where it is shown that the driven Liouville-von Neumann 
equation reproduces the Landauer results for those values of $\Gamma$ giving high currents.

\section{Application to organic polymers}

\begin{figure}
\begin{center}
\includegraphics[width=5.3cm]{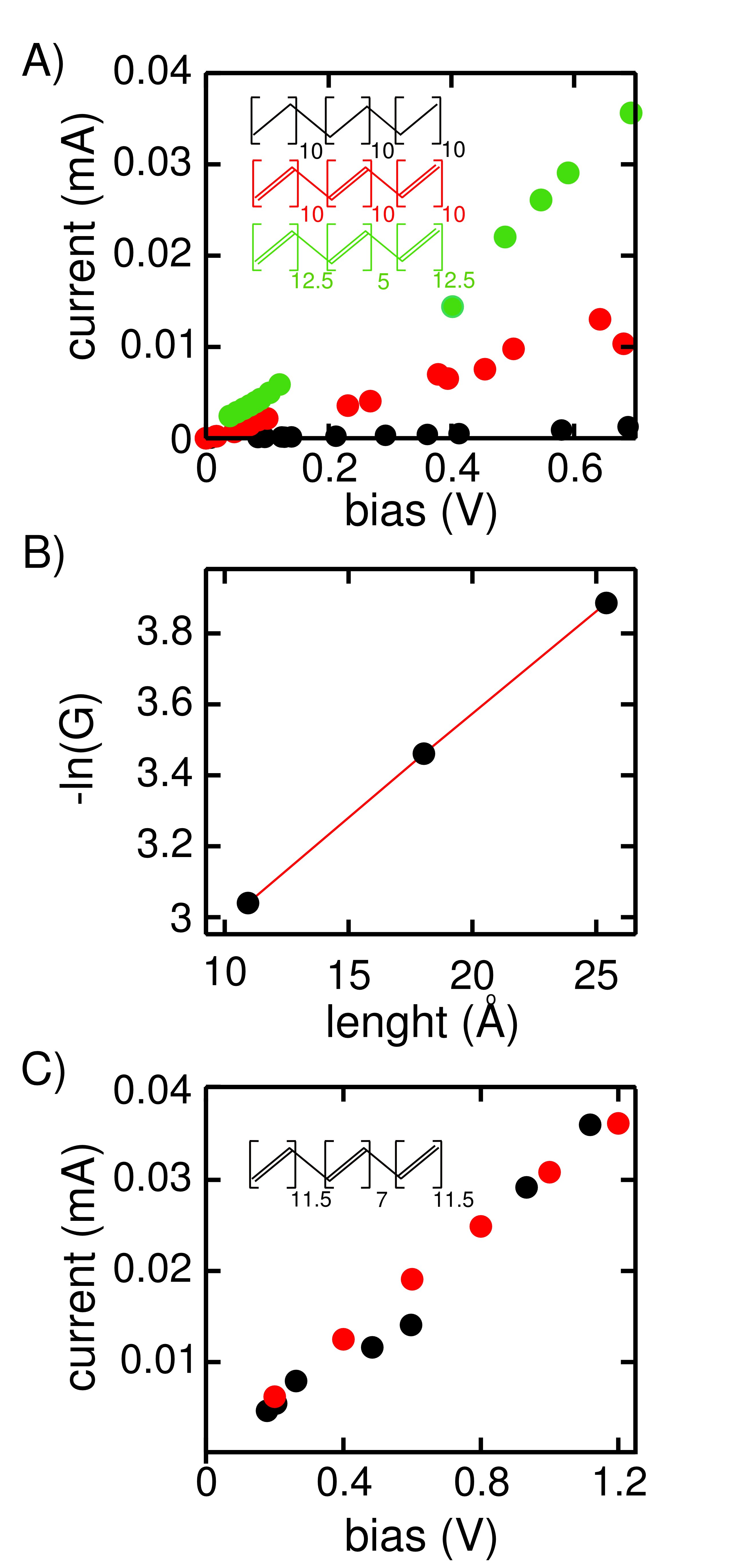}
\end{center}
\caption{A) $I-V$ curves computed with the driven Liouville-von Neumann approach
for different organic structures: 
a trans-polyacetylene chain with a bridge of 10 and 20 carbon atoms (green and
red dots respectively), and a 
saturated alkane of 60 carbon atoms  with 20 CH$_2$
units in the bridge (black dots).
B) Logarithm of the conductance obtained from the current-voltage characteristics, versus the distance between leads,
for the trans-polyacetylene model with bridges of  20, 14 and 10 carbon atoms.
C)  $I-V$ curves for a trans-polyacetylene chain of 14 carbon atoms. The red dots show results from NEGF calculations
for a model connected on each end to gold electrodes
through a sulfur atom (data extracted from reference \cite{superlattices}).
The black dots depict the steady-state currents computed through the driven Liouville-von Neumann approach,
using a Source and a Drain of 23 carbon atoms each.}
\label{figcomparative}
\end{figure}

Figure \ref{figcomparative}A compares the current-voltage characteristic of three hydrocarbons:
the polyacetylene structure with a bridge of 20 carbon atoms examined in the previous sections, 
the same molecule with a shorter bridge of 10 carbon atoms, and
a linear saturated alkane of 60 carbon atoms (CH$_3$-(CH$_2$)$_{58}$-CH$_3$) with 20 CH$_2$
units in the bridge.
In every  case, the calibration procedure based on the Mulliken charges difference
between Source and Drain, makes all data-points obtained with different $\Gamma$ values,
to collapse on a single curve. 
The currents for the saturated hydrocarbon are between one and two orders of magnitude
below those computed for the unsaturated molecule of the same size.
The conductance obtained for polyacetylene increases by a factor of 3 when the length
of the bridge is reduced from 20 to 10 carbon atoms. This result reflects a tunneling
decay constant ($\beta$) of 0.058 \AA$^{-1}$, in good agreement with the available
experimental estimates for such parameter in this polymer.\cite{CPHC-beta}
Figure \ref{figcomparative}B shows that, at least for the three different polyacetylene
lengths explored in this work, the method reproduces a perfectly exponential decay.

Panel C confronts, for a polyacetylene bridge of 14 carbon atoms, 
the currents obtained from our time-dependent simulations,
with those calculated through the NEGF formalism
using the TranSIESTA method.\cite{superlattices}
In the later calculations, the polyacetlylene chain is attached to two gold electrodes via sulfur atoms,
while in the real time simulations, the leads and the bridge have the same structure.
Noteworthy, and despite the distinct molecular junctions, the currents are in very good accord.
This suggests that the structure of the electrodes plays a minor role in the steady-state currents
obtained through the driven Liouville von Neuman approach.

\section{Summary}

This study presents the first implementation of the driven Liouville-von Neumann approach
for time-dependent transport
in an ab-initio DFT setting. The main modifications or innovations
with respect to previous semi-empirical schemes are:
(i) incorporation of charge injection in the off-diagonal elements of the driving operator;
(ii) preservation of the total charge through time-dependent absorbing and
injecting rate parameters; (iii) modulation of the perturbation associated with the
driving term at time zero; and (iv) adoption of the charge difference between leads as
the reference variable to establish the voltage bias, which removes the dependence
on the rate parameter.  

This has proved to be an efficient and stable scheme, suitable to perform real-time
electron transport simulations on systems above a hundred atoms for several hundreds of femtoseconds.
In the molecules explored here, steady states were typically achieved within the first ten or
twenty femtoseconds.
This methodology opens the door to simulations of charge
transport in realistic chemical structures,
from conducting polymers to metallic nanowires to biological macromolecules.
Aside from the most conventional phenomena, this method allows to explore a multiplicity of 
challenging and sophisticated conductance experiments, for example time-resolved transport modulated by
electric fields or laser pulses. In particular, although not discussed in the present
work, this code offers the possibility to represent large environments through a quantum-mechanics
molecular-mechanics approach.\cite{tddft} This can be useful to model the effect of a solvent
or other complex media in the transport process, which will be the subject of future work.

\section{Acknowledgment}

The present work is dedicated to the memory of Jorge Luis Morzan.

We thank Dario Estrin and Daniel Murgida for helpful discussions.
We express our gratitude to Ivan Girotto and the ICTP, for technical support and
GPU computing time which made possible most of the simulations presented in this study.
This work has been funded by grants
of the Agencia Nacional de Promocion Cientifica y Tecnologica
de Argentina, PICT 2012-2292,
and of The University of Buenos Aires, UBACYT 20020120100333BA.
UM and FR acknowledge CONICET for doctoral fellowships.

\newpage

\bibliography{references}

\end{document}